\documentclass{jpp}
\usepackage{graphicx}

\usepackage[colorlinks=true, linkcolor=blue, citecolor=blue]{hyperref}

\usepackage[utf8]{inputenc}
\usepackage[T1]{fontenc}
\usepackage{amsmath}

\shorttitle{Machine Learning-based models in particle-in-cell codes}
\shortauthor{C. Badiali, P. J. Bilbao, F. Cruz and L. O. Silva}

\title{Machine Learning-based models in particle-in-cell codes for advanced physics extensions}

\author{
  Chiara Badiali\aff{1}\corresp{\email{chiara.badiali@tecnico.ulisboa.pt}},
  Pablo J. Bilbao\aff{1},
  F\'{a}bio Cruz\aff{1, 2},
 \and Luis O. Silva\aff{1}\corresp{\email{luis.silva@tecnico.ulisboa.pt}}
}

\affiliation{\aff{1}GoLP/Instituto de Plasmas e Fus\~{a}o Nuclear, Instituto Superior T\'{e}cnico, Universidade de Lisboa, 1049-001 Lisbon, Portugal
\aff{2}Inductiva Research Labs}

\begin{document}

\maketitle

\begin{abstract}
In this paper we propose a methodology for the efficient implementation of Machine Learning (ML)-based methods in particle-in-cell (PIC) codes, with a focus on Monte-Carlo or statistical extensions to the PIC algorithm. The presented approach allows for neural networks to be developed in a Python environment, where advanced ML tools are readily available to proficiently train and test them. Those models are then efficiently deployed within highly-scalable and fully parallelized PIC simulations during runtime. We demonstrate this methodology with a proof-of-concept implementation within the PIC code OSIRIS, where a fully-connected neural network is used to replace a section of a Compton scattering module. We demonstrate that the ML-based method reproduces the results obtained with the conventional method and achieves better computational performance. These results offer a promising avenue for future applications of ML-based methods in PIC, particularly for physics extensions where an ML-based approach can provide a higher performance increase.
\end{abstract}

\section{Introduction}
\label{sec:introduction}

% Introduction to PIC
The use of computer simulations has had a major impact in furthering our understanding of kinetic plasma processes in a variety of laboratory and astrophysical scenarios. The particle-in-cell (PIC) method~\citep{dawson1983particle, hockney1988computer} has been the flagship technique for the study of these processes over the past decades. In the PIC method, plasma is described as a collection of charged particles interacting through self-consistent electromagnetic fields. In each simulation time step, the plasma charge and/or current densities are computed from the distribution of particles in configuration and velocity spaces and deposited on a grid that discretizes the simulation domain. These densities are then used to advance the electromagnetic field via Maxwell's equations, which in turn are used to advance particle momenta and positions.

% Standard PIC is scalable, but ever more complex modules are required to tackle challenging problems
The PIC method has found remarkable success in modelling a vast range of systems in plasma physics. Leveraged on the high scalability of PIC in high-performance computing (HPC) infrastructures, state-of-the-art codes routinely simulate $\sim 10^{10}$ particles in grids with $\sim 1000^3$ cells describing very large systems (\textit{i.e.}, with sizes well beyond plasma kinetic scales)~\citep{shukla2020interplay,arrowsmith2021generating}.

% Advanced applications require advanced methods, which need to be highly tailored and / or have large computational costs
The increasing complexity of the systems studied with the PIC algorithm has also led to the development of additional physics extensions to this method e.g. to model Coulomb collisions~\citep{takizuka1977binary, nanbu1997theory}, ionization~\citep{kemp2004modeling}, radiative losses~\citep{vranic2016classical} and Quantum Electrodynamics processes~\citep{vranic2016quantum}. However, these modules often include advanced numerical methods with significant computational overhead, hindering the performance and scalability of PIC. Another example of costly algorithms in PIC simulations is that introduced by numerical solvers that account for vacuum polarization~\citep{grismayer2021quantum} or particle equations in nontrivial spacetime metrics~\citep{bacchini2018generalized, parfrey2019first}, which require computationally expensive iterative methods or algorithms highly tailored to capture specific nonlinearities.

Advanced physics methods in PIC can introduce computational overhead for different reasons. In the specific case of binary collisions, particles are paired when close in configuration space and scattered according to a probability distribution function describing the collisional process ~\citep{takizuka1977binary, miller1994coulomb, sherlock2008monte, higginson2017full}. Depending on the complexity of the process, the probability may be computed, either analytically or numerically, or interpolated from large data tables stored in memory or in files, which can be highly computationally inefficient.

% Machine Learning methods can be used for data-driven discovery of tailoring techniques and compact representations relevant to PIC sub-modules
Machine Learning (ML) offers a promising avenue for the discovery of new algorithms that could facilitate the inclusion of advanced physics modules in PIC models via e.g. generalizable approximators of unknown functions or new solvers of partial-differential equations. Early works incorporating ML-based methods in the PIC loop focused on assisting~\citep{kube_2021} and replacing~\citep{aguilar2021deep} the numerical solver of Maxwell's equations, showing that its accuracy is preserved.

% Workflow to train Machine Learning models for PIC and to deployment them in production codes is not trivial
In spite of the potential of ML-based methods for PIC algorithms, developing and deploying ML-based methods in state-of-the-art PIC codes presents a substantial challenge. 
%\textcolor{red}{So far, the development and deployment of ML-based methods with both the core components and additional sub-modules of PIC codes remains largely unexplored.}
While the former are developed in modern languages such as Python or Julia due to their flexibility and large pool of open-source resources (e.g. the Python libraries TensorFlow~\citep{tensorflow_2015}, Keras~\citep{KERAS},  and PyTorch~\citep{pytorch_2019}), the latter are usually written in lower level languages such as Fortran or C++, to maximize performance and scalability in large HPC systems.

% We propose an ML-PIC workflow for OSIRIS
Integrating the ML model development and PIC production environments is essential to ensure an efficient model experimentation cycle. However, to the extent of our knowledge, this integration has not yet been discussed. In this work, we propose an interface to develop and deploy ML models in state-of-the-art PIC codes, in particular Neural Networks (NNs). This interface, based in open-source software, allows model training to be done in Python and final model parameters to be read and used for inference during a PIC simulation. We implement this interface in the PIC code OSIRIS ~\citep{OSIRIS, fonseca_2008} and demonstrate it in simulations using a Compton scattering module~\citep{compton_module} available in OSIRIS, demonstrating how to generalize ML techniques to address the challenges of advanced physics modules in the PIC algorithm.

% Paper outline
This paper is organized as follows: in Section~\ref{sec:ml_pic_interface}, a detailed description of the ML-PIC interface in presented. In Section~\ref{sec:ml_pic_poc}, we present a proof-of-concept use of that interface: the use case, concerning the evaluation of the probability of Compton scattering events, is detailed in Section~\ref{subsec:compton_in_pic}, the ML methods developed for this particular use case are presented in Section~\ref{subsec:ml_methods}, and a comprehensive study of the computational performance, accuracy and applicability of these methods in production PIC simulations is discussed in Section~\ref{subsec:results}. In Section~\ref{sec:conclusions}, our conclusions are presented.

\section{ML-PIC interface}
\label{sec:ml_pic_interface}

Recent works on assisting plasma kinetic simulations~\citep{aguilar2021deep, kube_2021} with ML-based methods offer promising prospects for efficiently solving challenging sub-steps of the PIC algorithm. It is thus expected that ML-based methods and plasma simulations will be used in tandem in the near future. Our work focuses on designing an efficient interface for the development and deployment of ML-based methods, in particular NNs, into state-of-the-art PIC codes. Such an interface is challenging to design, since ML-based models are usually developed in high-level environments, such as Python, while PIC codes tend to be written in lower-level languages such as Fortran. This duality of environments, although technically demanding to manage, allows users to leverage on the advantages of each of them.

Fortran is a fast and efficient computational language. Due to its high scalability in large HPC systems, it is widely used in large scientific computing applications. For these reasons, many PIC codes are written in Fortran, e.g. OSIRIS~\citep{OSIRIS, fonseca_2008}, EPOCH \citep{arber2015contemporary}, Tristan~\citep{buneman1993computer} and UPIC~\citep{decyk2007upic}. On the other hand, ML models are typically trained and run in high-level languages, such as Python, whose libraries focused on ML typically interface with code written in more efficient, lower-level languages (such as C/C++). This allows Python users to efficiently train and develop ML models. These libraries are also capable of running in parallel both during training and inference in CPU, GPU and other advanced architectures.

The lack of an interface between Fortran and Python environments makes the task of implementing ML-based models in highly scalable computational codes arduous. Recent efforts have significantly simplified this task. The micro-framework \texttt{neural-fortran}~\citep{curcic2019parallel} was developed with the intent of providing a library capable of performing the basic operations needed for a NN (e.g. gradient descent optimization, fully-connected dense layer calculations) within a Fortran environment. Further work expanded this library into the Fortran-Keras Bridge (FKB) \citep{ott2020fortran}. With FKB, one can train NNs in a Python environment with the Keras library and export their parameters to files that are then read and used for inference by the FKB Fortran. This library constitutes one of the first attempts at bridging both environments and facilitates the use of NNs in highly scalable, computationally intensive Fortran codes.

\begin{figure}
    \centering
    \includegraphics[width=3.3in]{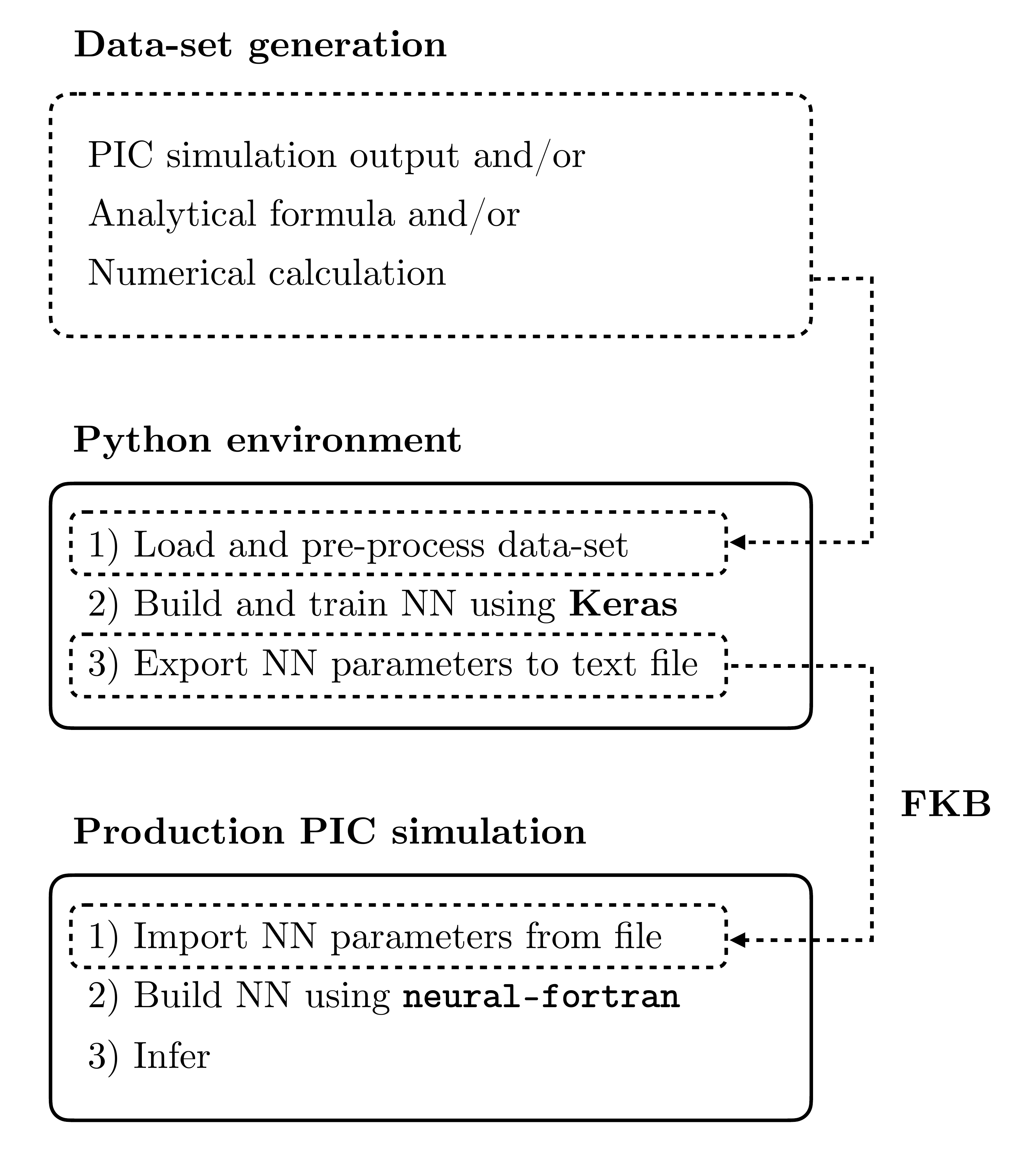}
    \caption{Schematic representation of the data acquisition, ML model training and production workflow. The ML model is loaded in production simulations during run-time using the Fortran-Keras Bridge (FKB) and interpreted with the \texttt{neural-fortran} micro-framework within PIC.}
    \label{fig:ml_pic_scheme}
\end{figure}

In this work, \texttt{neural-fortran} and the FKB library are used in the OSIRIS code to run ML-based models in production PIC simulations. In Fig. \ref{fig:ml_pic_scheme} we outline the workflow necessary to efficiently train and utilize a NN model within OSIRIS. Firstly, a training data-set must be obtained. 
Our data was generated using algorithms previously implemented in OSIRIS and was later loaded into the Python environment and employed to train a NN.

After the training has been performed, the FKB Python library is used to export the model into a compatible format for the Fortran side of the workflow. OSIRIS has been modified to include a NN object which stores the model into memory, from the exported file, at the beginning of the simulation. Moreover, this NN object has a subroutine which can run the model and execute it for any given batch of inputs. FKB originally included a version of \texttt{neural-fortran}. Thus, FKB is capable of performing the same operations as \texttt{neural-fortran}. Over time, the libraries have diverged and \texttt{neural-fortran} has continued to be further optimized. For this reason the initial loading is done with the FKB Fortran implementation and then \texttt{neural-fortran} is employed in the subroutine responsible for the model computation. During runtime, after the model has been loaded into memory, the the NN object can be invoked and given inputs in the same manner one would call any other subroutine.

As far as we are aware, the workflow outlined here is the first of its kind for PIC simulations. In the next section, a NN is developed and deployed to replace a long analytical calculation in PIC simulations. A direct comparison between the computational performance of the NN inference and the analytical calculation in a production environment was performed, to demonstrate the potential advantages of ML-based techniques in the context of production PIC runs.

\section{ML-PIC proof of concept}
\label{sec:ml_pic_poc}

As a proof of concept, here a NN that replaces an analytical calculation in a Compton scattering module for PIC is presented. The choice of this specific task has been driven by its simplicity, which allows us to have better control of the performance of the classical and ML-based methods.

\subsection{Compton scattering in PIC}
\label{subsec:compton_in_pic}

Compton scattering describes the most basic interaction of radiation with matter via the binary scattering between leptons and photons~\citep{compton1923quantum}. Monte Carlo techniques are usually employed to implement this phenomenon in PIC simulations,  as the quantum nature of the process is intrinsically stochastic. Here, we quickly outline the current implementation of the Compton scattering module in the OSIRIS PIC code. For an in-depth description of this implementation, we refer the readers to~\citet{compton_module}.

The algorithm follows three steps. Firstly, the simulation space is divided into collision cells, in which the particles are binned. From this binning, a list of colliding leptons and photons is produced for each cell. Secondly, the No-Time-Counter (NTC) method is employed to significantly shorten the list of pairs to be collided~\citep{bird1989perception}. For each pair in this list, the probability of interaction is then calculated and compared against a random number. Finally, for each pair deemed to collide, the kinetic collision is resolved and the momentum of the particles is updated.

\subsection{ML methods}
\label{subsec:ml_methods}

Our proof of concept aims at replacing the calculation of the probability of interaction via Compton scattering. This probability is a function of the weights of the macro-particles, their momentum and a normalization factor. This calculation is detailed in Sec. 3 of ~\citep{compton_module}. The photon momentum is first Lorentz boosted to the lepton frame. Then, the probability is calculated using the Klein-Nishina cross-section and conservation of momentum. Finally, the evaluated cross-section is Lorentz boosted back into the simulation (lab) frame to obtain the probability of interaction between time steps. This calculation is a good candidate for an ML proof-of-concept, mainly because, being analytical and optimized within the PIC, it represents the worst case testing scenario for an ML model to compete against. A NN is used to predict a value a value $y' \in [0, 1]$, corresponding to the probability of interaction. The NN takes as input the momenta of the macro-photons and macro-leptons in the simulation frame in units of $m_e c$ ($\mathbf{p}_\gamma$ and $\mathbf{p}_e$, respectively) and the maximum probability of interaction in each PIC cell $P_{\max} = 2 \sigma_T c \Delta t \max{\left[ w_e , w_\gamma \right]}$, where $\sigma_T$ is the Thompson cross-section, $\Delta t$ is the time step and $\max{\left[ w_e , w_\gamma \right]}$ is the maximum macro-particle weight within the cell~\citep{compton_module}. During the training process, the analytical probability of interaction $y$ is used to compute a loss function and to update the NN via back-propagation.
 
The training data was collected from different 2D3V OSIRIS simulations. In these simulations, a uniform electron-positron-photon plasma with isotropic waterbag distributions in momentum space ($f(p) = n_0 H(p+p_0)$, where $n_0$ is a constant density, $H(p)$ is the Heaviside function and $p = | \mathbf{p} |$ is the magnitude of the particle momentum) was initialized and allowed to interact through Compton scattering. Every time a probability calculation was performed, the result was outputted alongside the relevant information employed in the calculation. Once the simulation was finished, all the data was collected for training. Three data-sets were produced from three simulations with $p_0 / m_e c = 25$, $50$ and $100$.

\begin{figure}
    \hspace*{\fill}
    \raisebox{-1.1\height}{\includegraphics[width=0.35\linewidth]{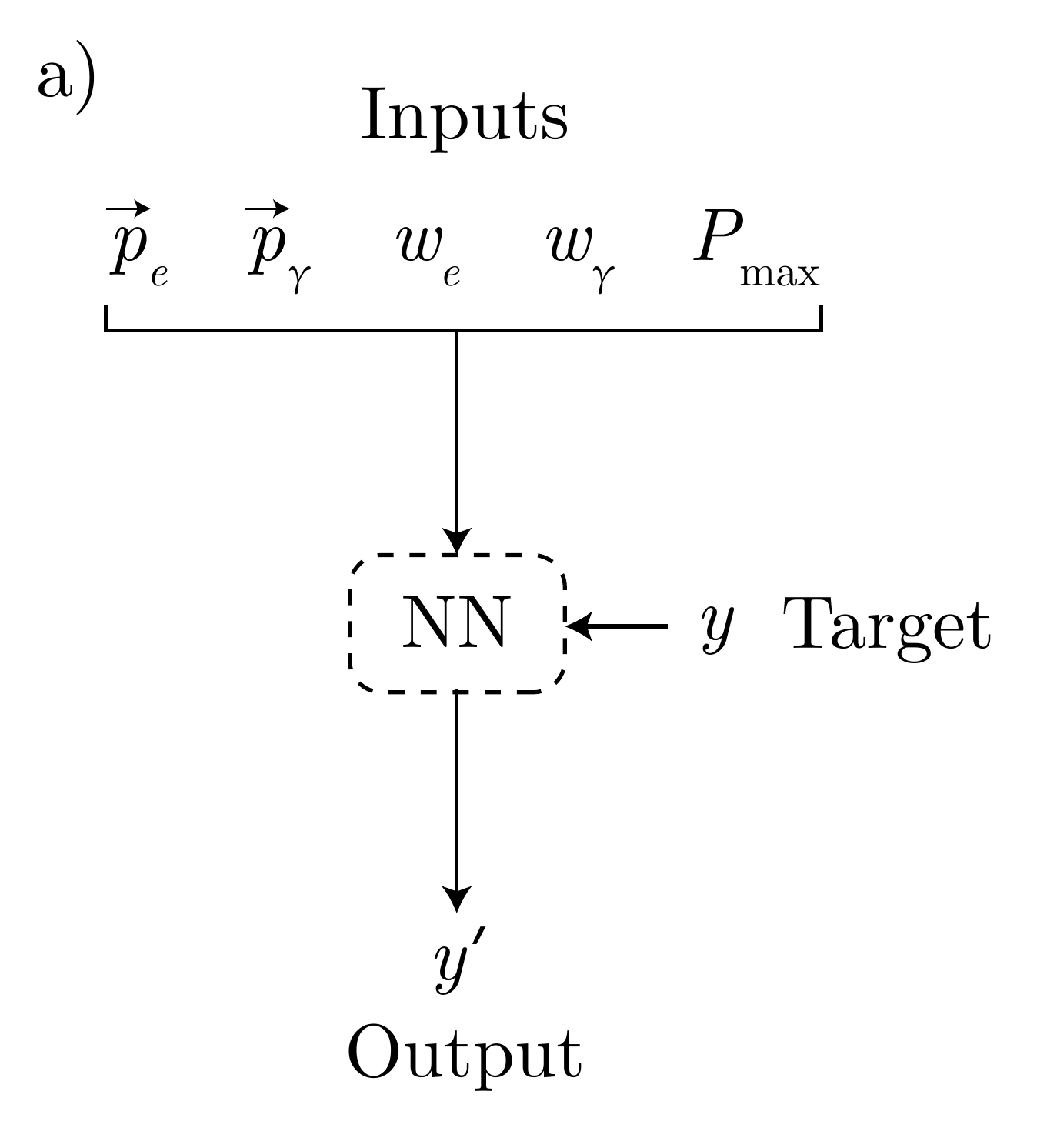}}
    \hfill
    \raisebox{-\height}{\includegraphics[width=0.6\linewidth]{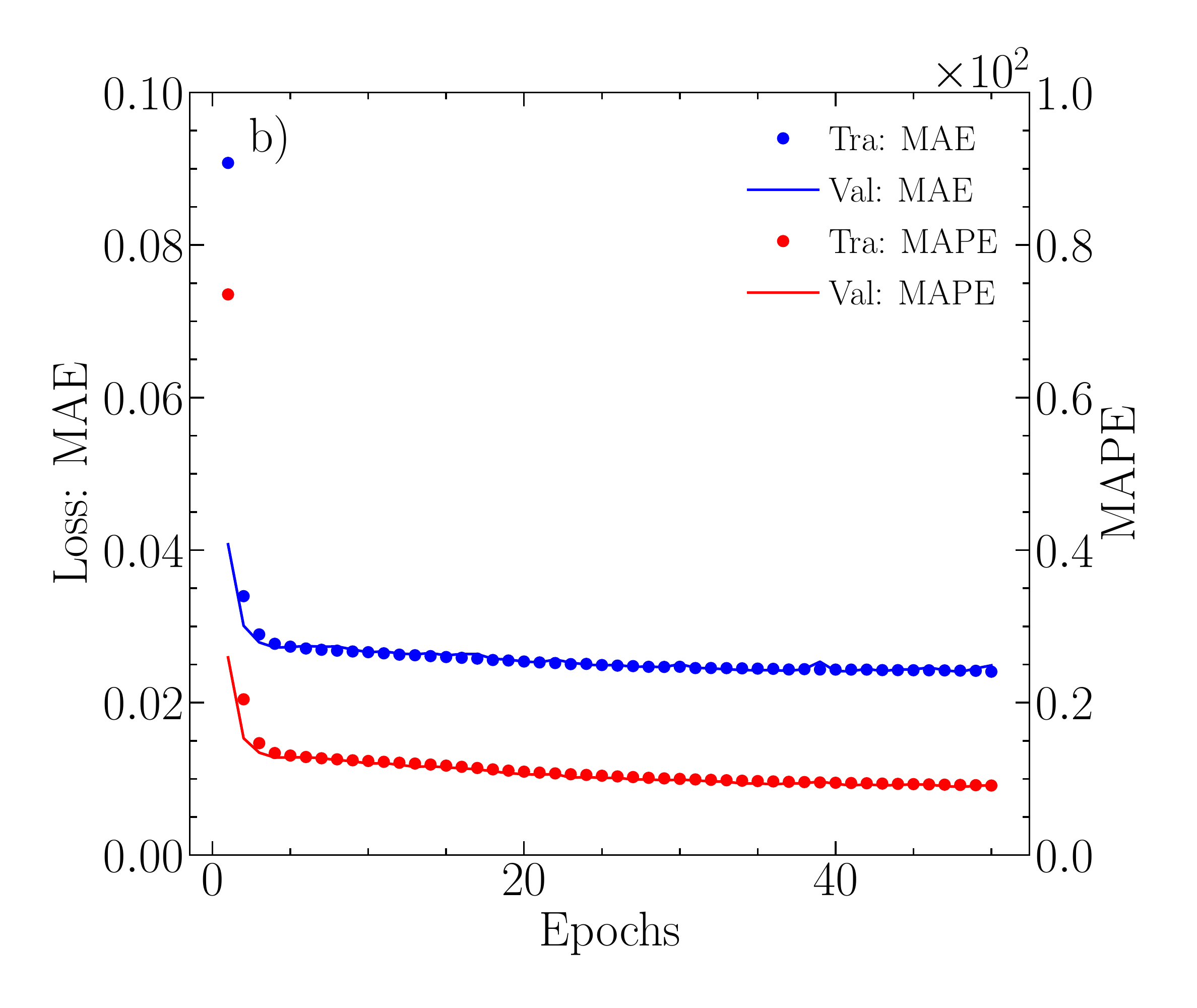}}
    \hspace*{\fill}

    \caption{a) Depiction of the inputs of the neural network during the training process: $\mathbf{p}_e$ and $\mathbf{p}_\gamma$ are the three momentum of the macro-particles undergoing collisional process, $w_e$ and $w_\gamma$ are their numerical weights and $P_{\max}$ is the maximum probability of interaction in each PIC cell. Finally, the target $y$ constitutes the probability of interaction according to an analytical model and $y'$ is the prediction of the neural network. b) Loss score, \textit{i.e.}, the mean absolute error (MAE) and Mean Absolute Percentage Error (MAPE) as accuracy for training and validation data as a function of the training epoch for the training data with $p_0 = 25\ m_e c$.}
    \label{fig:struct_loss}
\end{figure}

The NN architecture was chosen such that each output computation was computationally inexpensive. We used a small number of layers and neurons per layer, which translated into a small number of floating point operations. By keeping the neural layers as simple as possible (\textit{i.e.}, only employing fully-connected layers), we ensure that all operations can be vectorized in all common compilers both in high- and low-level languages. Thus, the NN structure consisted of 3 hidden layers with 9, 12, 9 neurons each and 1 neuron on the output layer. The hidden layers used ReLU~\citep{fukushima1969visual} as activation functions, whereas the output layers used sigmoid~\citep{han1995influence}, to guarantee that the predicted value was in the range $[0, 1]$. As seen in Fig.~\ref{fig:struct_loss}a, the NN was trained by providing the relevant information of many photon-lepton pairs and their associated probability of interaction. The NN computes the probability of interaction between the macro-particles $y'$. During the training process, many targets and outputs of the NN are compared employing a loss function~\citep{Chollet}, that we took to be the Mean Absolute Error MAE  $=(\sum_{i= 1}^n|{y'_i-y_i|})/n$, where $n$ is the sample size, $y'_i$ is the $i$-th predicted value and $y_i$ is $i$-th true value. The loss function characterizes the inaccuracy of the model prediction, and is used to repeatedly update the NN parameters (in our case, using the Adam optimizer with a learning rate $r = 0.001$ ~\citep{kingma2014adam}). As more training data is fed to the NN, it learns the target function and the loss score is minimized. A total of 5000 epochs were used in the training. The results of the training process for the data-set with $p_0 = 25\ m_e c$ are shown in Fig.~\ref{fig:struct_loss}b for both the training and the validation data-sets (corresponding respectively to a randomly selected $80\%$ and $20\%$ fraction of the total data, that comprised $\approx 10^7$ Compton scattering events).

To achieve an efficient training process, two techniques were employed. First, the training data was balanced ~\citep{DiBello2021}, so that Compton scattering events with different probabilities appeared a similar amount of times. This is a critical step for efficiently learning the probability of interaction via Compton scattering, as it quickly decays with increasing photon momenta in the lepton frame. Second, we employed the multi-stage training process Transfer Learning (TL)~\citep{transfer}. This technique trains the NN in stages corresponding to a different subsets of the input parameter space $S$, allowing it to train on the full probability distribution while relaxing the condition that the training data has to be identical in all subsets of $S$.

We used three stages of TL. In the first stage, the NN was trained with the data-set containing samples of low energy interactions between macro-leptons and macro-photons ($p_0 / m_e c = 25$), corresponding to high probabilities of interaction. In the second and third stages, the NN was smoothly introduced to higher values of particle energies (and hence smaller probabilities of interaction), through data-sets containing of high ($p_0 / m_e c = 50$) and very high ($p_0 / m_e c = 100$) energy interactions, respectively. In each TL stage, the particle energy range includes also that of previous TL stages in order to avoid \textit{catastrophic forgetting}~\citep{kirkpatrick2017overcoming}, \textit{i.e.}, having the NN lose information about the previous training stage while focusing on the relevant information of the current task.

\subsection{Results}
\label{subsec:results}

In order to assess the value of our ML-assisted Compton scattering module for PIC codes, we studied: i) its accuracy, \textit{i.e.} its ability to correctly capture the physical properties of the collisional process and produce results that match analytical approaches, and ii) its computational performance, to determine whether this method yields a speed up relative to the implementation described in section~\ref{subsec:compton_in_pic} (and in Ref.~\citep{compton_module} in more detail). 

\begin{figure}
    \centering
    \includegraphics[width=\linewidth]{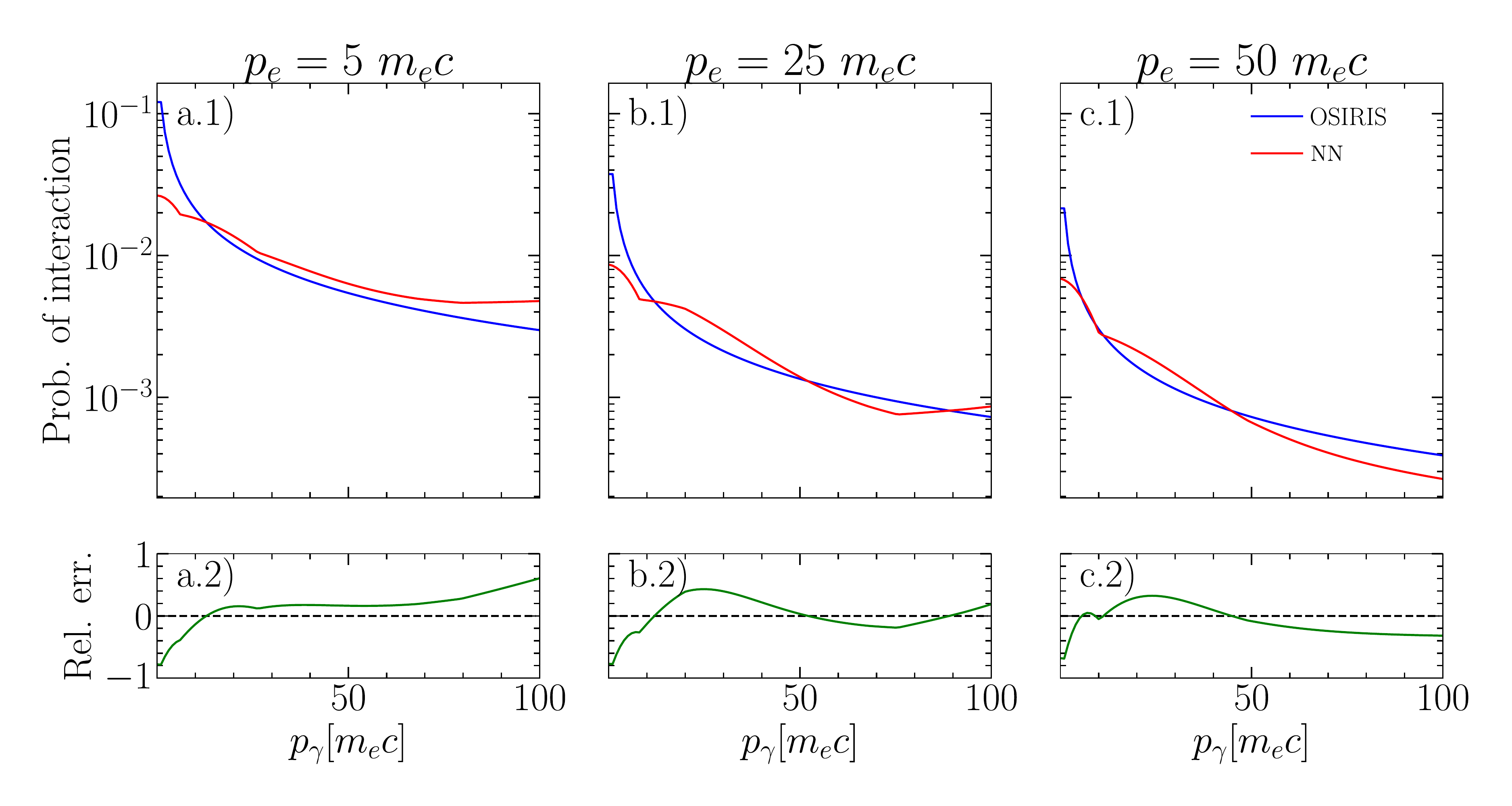}
    \caption{Probability of interaction estimated for the No-Time-Counter method as a function of the incoming photon energy colliding head on with an electron with momentum $p_e = 5$, $25$ and $50~m_e c$ (a-c1, respectively). The estimate calculated by OSIRIS (blue) and the prediction by neural network module for OSIRIS (red) are directly compared and their relative error is shown in a-c2.}
    \label{fig:cross_comp}
\end{figure}

The first benchmark corresponded to a comparison between the prediction of the NN and the analytical probability of interaction. This test aimed at explicitly showcasing how ML-based methods can learn and replicate the Compton scattering process. In Fig.~\ref{fig:cross_comp}a-c, we show the scattering probability between electrons with momenta $p_e / m_e c = 5$, $25$ and $50$ and photons with a wide range of momenta $p_\gamma$. The good agreement between the two methods results in relative errors smaller than order unity for all tested $p_\gamma$. The NN performs worse in the region of low $p_{\gamma}$ because training was focused in higher energy collisions, with each step in the TL scheme moving further into higher energies. To improve this result, another TL training could be done in the region of low photon energy.

To further test the accuracy of our method, we performed a benchmark simulation consisting of an electron beam interacting with a monoenergetic photon gas via inverse Compton scattering~\citep{blumenthal1970bremsstrahlung}, which has also been performed with the conventional implementation of Compton scattering in OSIRIS presented in a previous work~\citep{compton_module}.

\begin{figure}
    \centering
    \includegraphics[width=\linewidth]{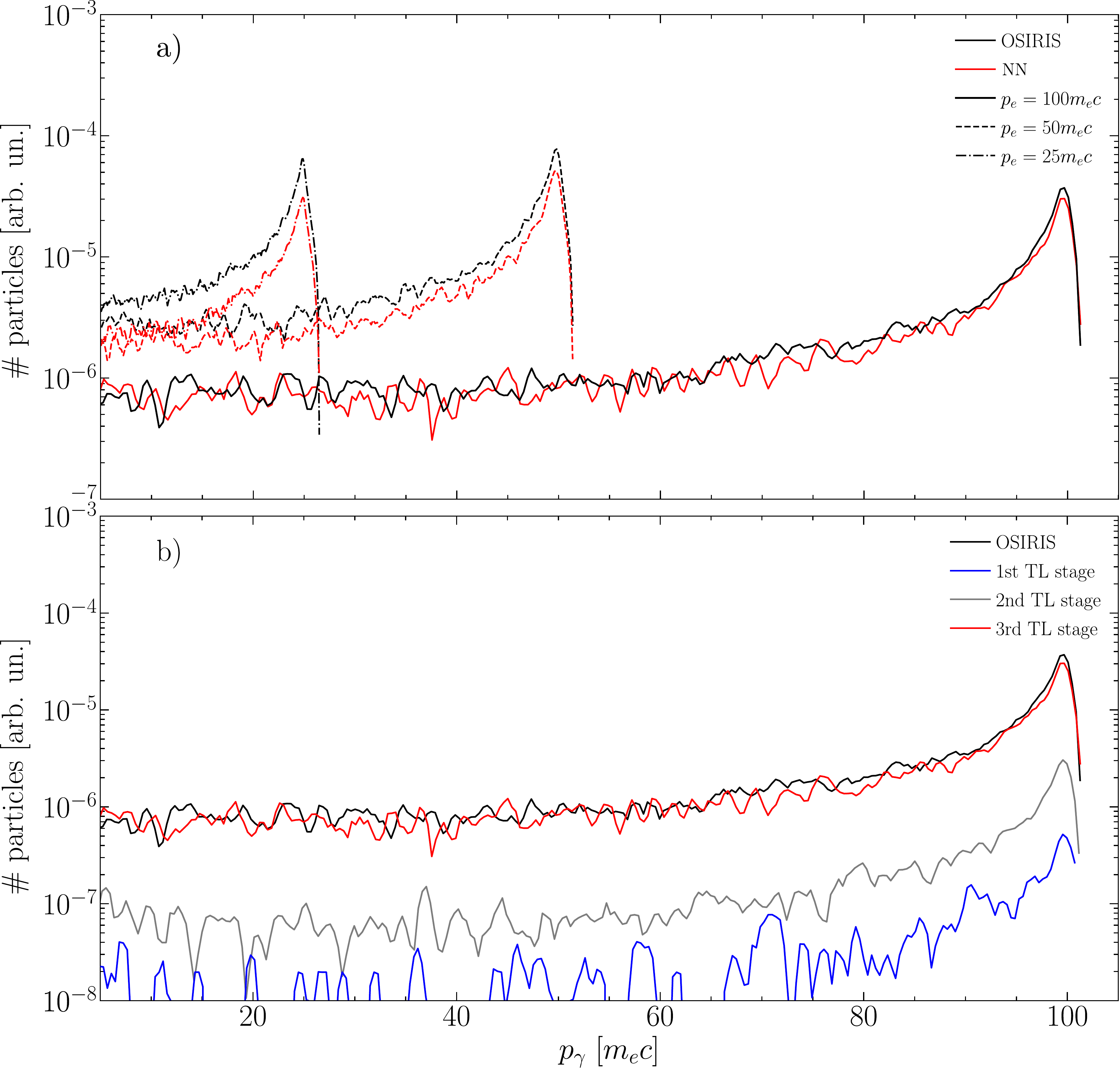}
    \caption{a) Scattered photon momentum distribution resulting from inverse Compton scattering with an electron beam with momentum $p_e$. The three peaks correspond to three initial monoenergetic electron beams with $p_e = 25$, $50$ and $100\ m_e c$ (identified with dash-dotted, dashed and solid lines, respectively) interacting with a monoenergetic photon gas with energy $2.5 m_e c^2$. For each photon beam energy, the neural network module for OSIRIS prediction (red) is plotted against the results from the conventional algorithm (black).
    b) Scattered photon momentum distribution from inverse Compton scattering between an electron beam with $p_e / m_e c = 100$ and a mono-energetic photon gas with energy $\varepsilon_\gamma / m_e c^2  = p_\gamma / m_e c = 2.5$. Different model predictions at different Transfer Learning (TL) stages are plotted (blue, grey and red) against results obtained with the conventional Compton scattering algorithm in OSIRIS (black).
    }
    \label{fig:spec_peaks}
\end{figure}

In Fig.~\ref{fig:spec_peaks}, we show the photon spectra obtained from simulating the interaction of an electron beam of momentum $p_e / m_e c = 25$, $50$ and $100$ with an initial mono-energetic photon distribution of energy $\varepsilon_\gamma / m_e c^2 = 2.5$ using both the analytical and NN inference of the probability of interaction via Compton scattering. In all simulations (different $p_e$, analytical or NN), we observe a peaked photon momentum distribution developing during the interaction around $p_\gamma \approx p_e$, with an extended tail for $p_\gamma < p_e$. The NN and the analytical result agree within $\sim 10^{-5}$ particle count, \textit{i.e}, the NN precisely modelled the expected photon spectra for all the three peaks. A small underestimation of the number of collisions is observed for the peaks corresponding to $p_e/m_e c = 25$ and $50$, which is a consequence of the underestimation of the probability of interaction for small $p_\gamma$ identified above. Overall, our results show that Compton scattering physics can be accurately described by a NN, and thus that it can be used to replace conventional analytical calculations of the probability of interaction.

In order to illustrate the relevance of TL in the training process, we show in Fig.~\ref{fig:spec_peaks} the scattered photon momentum distributions for a simulation with $p_e / m_e c = 100$ obtained with NNs trained with a different number of TL stages. We can see a clear convergence toward the results obtained with the analytical approach with increasing number of TL stages. These results suggest that simulations can be performed with NNs trained in an already expected electron and/or photon energy range, and extra TL stages can be applied if at some point particle energies exceed the training range. Besides the improvement in accuracy, applying extra TL stages has the advantage of requiring a small number of training samples and thus does not significantly affect the cost of training.

%After checking the accuracy of the method, we can benchmark the computational performance of our ML-based method.
\begin{figure}
    \centering
    \includegraphics[width=\linewidth]{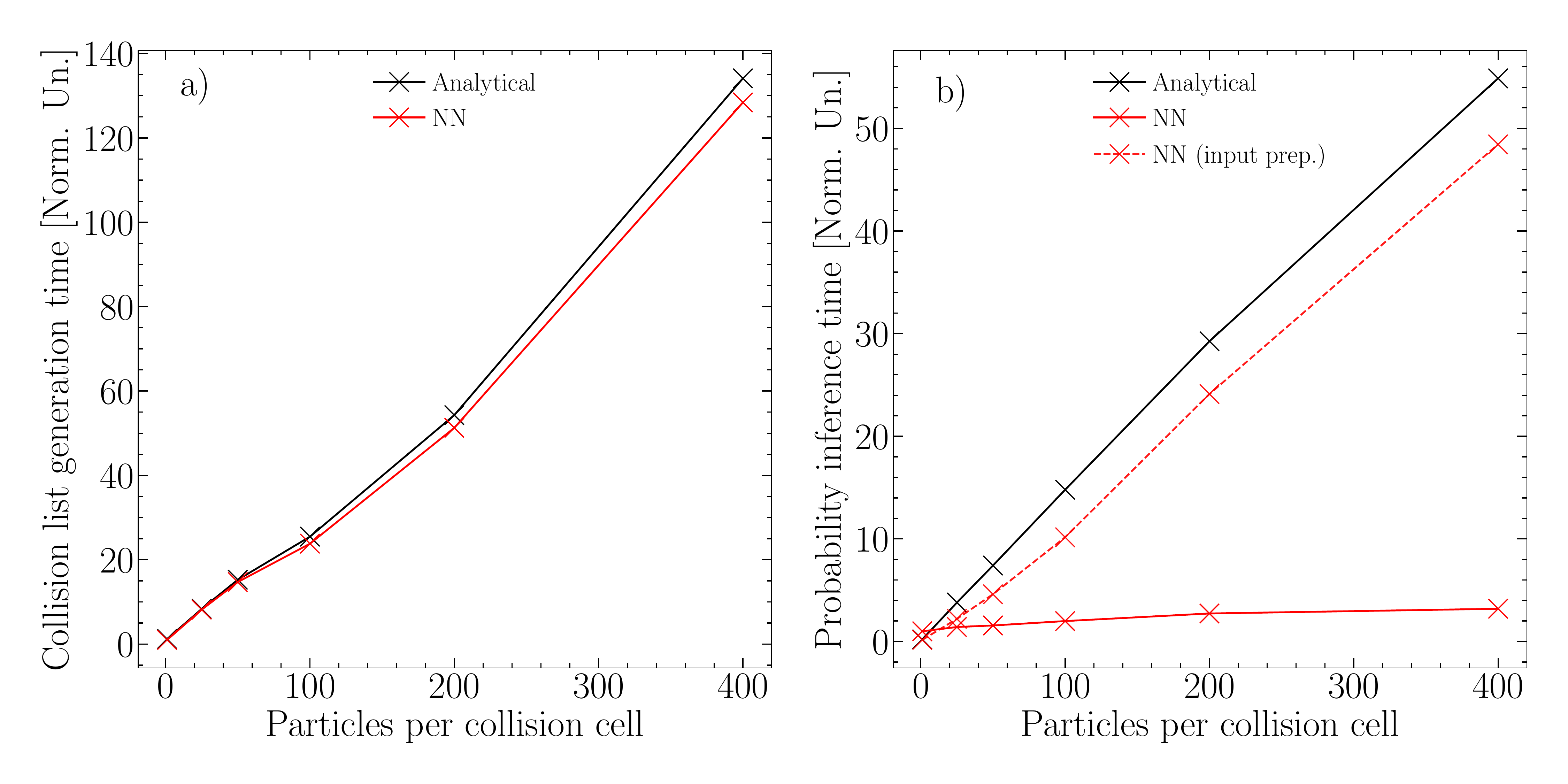}
    \caption{a) Comparison of the elapsed time during the complete collision list generation algorithm. b) Comparison of the elapsed time for the inference of the collision probability and for the NN case the input preparation is also plotted. The time is normalised with respect to the time taken for the conventional approach with $\mathrm{ppc}=1$.}
    \label{fig:performance_results}
\end{figure}

We have also benchmarked the computational performance of our ML-based method. Several simulations with varying number of particles per collision cell were performed and timed. We chose to benchmark our ML-based method against the original algorithm in OSIRIS, which is heavily optimized. The direct comparison between the two methods is plotted in Fig.~\ref{fig:performance_results}. In order to guarantee that both methods were compared on the same footing, we timed the computation of the probability of interaction with the NN and with the analytical approach within the same simulation. In each time iteration of these simulations, the analytical probability of interaction was used to determine if each pair of macro-particles should be collided. This ensured that both methods compute the probability for the same number of collisions in each time step, and that they are provided exactly the same inputs in each computation.

The performance comparison in Fig.~\ref{fig:performance_results}.a displays the total collision list generation time for the analytical and NN approach. It demonstrates that both methods scale linearly with respect to the number of particles per collision cell. The NN based approach outperforms the optimized Compton module, which becomes more evident with higher number of particles. The whole NN algorithm achieves only slightly faster times than the analytical based collision list algorithm. The reason behind this is that, even if the NN inference calculation is significantly faster than its analytical counterpart, as seen in Fig.~\ref{fig:performance_results}.b, the NN algorithm requires that the input of the NN is prepared. This part of the algorithm slows down the whole collision list generation.

\section{Conclusions}
\label{sec:conclusions}

In this work, we have proposed an interface to include ML-based methods in production PIC simulations. This interface allows NNs to be trained using open-source Python packages and deployed in PIC codes written in Fortran. We have presented a proof-of-concept implementation of this interface in the OSIRIS PIC code, and used it to perform simulations where an involved analytical calculation is replaced with a NN of parameters determined in a Python environment. The NN, which was trained with analytical data, computes the probability for two macro-particles to interact via Compton scattering. In order to ensure that the NN accurately captures the dynamic range of the target probability, we have i) employed multiple stages of TL corresponding to different particle energy ranges and ii) used a balanced data-set within each stage. We have tested the performance of our approach by studying the photon spectra produced in the interaction between an electron beam and a monoenergetic photon distribution. The ML-based method accurately reproduced the results obtained with a conventional algorithm using the analytical approach. We have also demonstrated that the ML-based method was able to outperform the analytical model in computational efficiency.

In the simulations presented in this work, the NN was trained in a Python environment and its parameters were then saved for later run-time use.
However, our implementation including \textit{neural-fortran} also allows training NNs within the PIC simulation and exporting them for use in other environments or other PIC simulations. This flexible architecture offers a promising avenue for future applications of ML-based methods in PIC, not only limited to collisional processes (e.g. Compton scattering, Coulomb collisions), but also in a vast range of physical processes (e.g. high-order QED processes) and numerical techniques, such as field and particle equation solvers, dynamical load balancing and advanced diagnostics.

\vspace{1cm}
C. Badiali and P. J. Bilbao contributed equally to this work. We would like to thank E. P. Alves and R. A. Fonseca for fruitful discussions, and for their comments on the manuscript. This work was supported by the European Research Council (ERC-2015-AdG Grant 695088). C. Badiali was also supported by FCT (Portugal) (grant PRT/BD/152270/2021). P. J. Bilbao was also supported by FCT (Portugal) (grant UI/BD/151559/2021). F. Cruz was also supported by FCT (Portugal) (grant PD/BD/114307/2016) in the framework of the Advanced Program in Plasma Science and Engineering (APPLAuSE, FCT grant PD/00505/2012). All simulations and tests were performed in the IST cluster, Lisbon. We graciously thank the comments from the anonymous referee for detailed and insightful review of our work, including suggestions that significantly improved the quality of the results, as well as of the manuscript. \\ 
\\ Declaration of Interests. The authors report no conflict of interest.

\bibliographystyle{jpp}

\bibliography{jpp-instructions}

\end{document}